\newcommand{\Le}{\ell}

\documentclass[aps,prb,twocolumn,eqsecnum]{revtex4-1}
\usepackage{amsmath,amssymb,amsfonts,mathptmx,graphicx,bm,color}

\begin{document}

\title{Variations on the adiabatic invariance: the Lorentz pendulum}

\author{Luis~L.~S\'{a}nchez-Soto and Jes\'{u}s Zoido}

\affiliation{Departamento de \'Optica, Facultad de F\'{\i}sica,
  Universidad Complutense, 28040~Madrid, Spain}
 
\date{\today}

\begin{abstract}
  We analyze a very simple variant of the Lorentz pendulum, in which
  the length is varied exponentially, instead of uniformly, as it is
  assumed in the standard case. We establish quantitative criteria for
  the condition of adiabatic changes in both pendula and  put in
  evidence their substantially different physical behavior with regard
  to adiabatic invariance.

\end{abstract}

\maketitle

\section{Introduction}

As early as 1902, Lord Rayleigh~\cite{Rayleigh:1902lr} investigated a
pendulum the length of which was being altered uniformly, but very
slowly (some mathematical aspects of the problem were treated in 1895
by Le Cornu~\cite{LeCornu:1895lr}). He showed that if $E(t)$ denotes
the energy and $\Le(t)$ the length [or, equivalently, the frequency
$\nu (t)$] at a time $t$, then
\begin{equation}
  \label{eq:adinv1}
  \frac{E (t)}{\nu (t)} =  \frac{E (0)}{\nu (0)} \, .
\end{equation}
This expression is probably the first explicit example of an adiabatic
invariant; i.e., a conservation law that only holds when the
parameters of the system are varied very slowly. The name was coined
by analogy with thermodynamics, where adiabatic processes are those
that occur sufficiently gently.

At the first Solvay Conference in 1911, Lorentz, unaware of Rayleigh's
previous work, raised  the question of the behavior of a
``quantum pendulum'' the length of which is gradually
altered~\cite{Langevin:1912fj} (by historical
vicissitudes~\cite{Navarro:2006qe} his name has become inextricably
linked to that system).  Einstein's reply  was that ``if the length of
the pendulum is changed infinitely slowly, its energy remains equal to
$h \nu$ if it is originally $h \nu$'', although no detail of his
analysis are given. In the same discussion, Warburg insisted that the
length of the pendulum must be altered slowly, but not systematically.
As Arnold aptly remarks~\cite{Arnold:1978yq} ``the person changing the
parameters of the system must no see what state the system is in.
Giving this definition a rigorous mathematical meaning is a very
delicate and as yet unsolved problem. Fortunately, we can get along
with a surrogate. The assumption of ignorance of the internal state of
the system on the part of the person controlling the parameter may be
replaced by the requirement that the change of parameter must be
smooth; i.e., twice continuously differentiable''.  This important
point is ignored in most expositions of adiabatic invariance.

Ehrenfest, who did not attend the Solvay Conference and was not
cognizant of that discussion, had indeed read Rayleigh's paper and
employed those ideas to enunciate his famous adiabatic
principle~\cite{Ehrenfest:1916cr}, which was promptly reformulated by
Born and Fock~\cite{Born:1928uq} in the form we now call adiabatic
theorem~\cite{Kato:1950kx}.  In fact, this was a topic of uttermost
importance in the old quantum
theory~\cite{Sommerfeld:1919qy,Born:1925uq}. To put it simply, if a
physical quantity is going to make ``all or nothing quantum jumps'',
it should make no jump at all if the system is perturbed gently, and
therefore any quantized quantity should be an adiabatic invariant. The
reader is referred to the book of Jammer~\cite{Jammer:1966ys} for a
masterful review of these questions, as well as the lucid and concise
mathematical viewpoint of
Levi-Civitta~\cite{Levi-Civita:1928tg,Levi-Civita:1934fr}.

The topic of adiabatic invariance has undergone a resurgence of
interest from various different fields, such as plasma physics,
thermonuclear research or geophysics~\cite{Whiteman:1977uq}, although
perhaps Berry's work on geometric phases~\cite{phases:1989oq} has put
it again in the spotlight. The two monographs by Sagdeev \textit{et
  al}~\cite{Sagdeev:1988ve} and Lochak and
Meunier~\cite{Lochak:1988bh} reflect this revival. More recently, the
issue has renewed its importance in the context of quantum control
(for example, concerning adiabatic passage between atomic energy 
levels~\cite{Oreg:1984fk,Gaubatz:1988kx,Gaubatz:1990ys,
Schiemann:1993zr,Pillet:1993ly,Kral:2007bh}),
as well as adiabatic quantum 
computation~\cite{Farhi:2000qf,Farhi:2001cr,Pachos:2001nx}.

Adiabatic invariants are presented in most textbooks in terms of
action-angle variables, which involves a significant level of
sophistication.  Even if a number of pedagogical papers has tried to
alleviate these difficulties~\cite{Parker:1971gf,Calkin:1977kx, 
Gignoux:1989fj,Crawford:1990rr,Anderson:1992uq,Pinto:2000bh,
Wells:2007ys,Shore:2009cr}, students often understand this notion
only at a superficial level. Actually, perfunctory application of the
adiabaticity condition may lead to controversial conclusions, even in
the hands of experienced practitioners~\cite{Marzlin:2004uq,Sarandy:2004zr,Du:2008kx,Amin:2009vn}.

Quite often the Lorentz pendulum is taken as a typical example to
bring up this twist for graduate
students~\cite{Wickramasinghe:2005yq,Kavanaugh:2005rt}. In spite of
its apparent simplicity, the proof of invariance is genuinely
difficult~\cite{Krutkow:1923dq,Kulsrud:1957fk,Gardner:1959uq,
  Kruskal:1962fj,Littlewood:1963lr,Brearley:1966ly,Werner:1969mz,Ross:1979kl}
and details are omitted.  The purpose of this paper is to re-elaborate
on this topic, putting forth pertinent physical discussion that
emphasizes the motivation for doing what is done, as well as to
present some variation of the Lorentz pendulum in the hope that its
solution will shed light on the subject at an intermediate level.

\section{The uniformly varying pendulum}

\subsection{Basic equations of motion}

We confine our attention to the ideal case of a simple pendulum of mass
$m$ and variable length $\Le (t)$, oscillating under the gravity. The
Lagrangian of the system is
\begin{equation}
  \label{eq:Laglet}
  L = \frac{1}{2} m \left [ 
    \left ( \frac{d \Le}{d t} \right )^{2}  +
    \Le^{2}  \left ( \frac{d \vartheta}{d t} \right )^{2} 
  \right ] 
  +  m g \Le \cos \vartheta \, ,
\end{equation}
where $\vartheta (t)$ denotes the inclination of the pendulum with the
vertical. The Euler-Lagrange equation for the generalized coordinate
$\vartheta$ becomes
\begin{equation}
  \label{eq:eqmotnl}
  \frac{d^{2} \vartheta}{d t^{2}} + 
  \frac{2}{\Le}  \left ( \frac{d \Le}{d t} \right )  \frac{d
    \vartheta}{d t}  +
  \frac{g}{\Le} \sin \vartheta = 0 \, .
\end{equation}
Note in passing that the length $\Le (t)$ acts as a geometrical (or
holonomic) constraint, which here becomes time
dependent~\cite{Goldstein:1980fk,Jose:1998ve}.

In what follows we shall restrict ourselves to the regime of small
oscillations (that is, $\sin \vartheta \simeq \vartheta$). In this
Section, we deal with the example of a pendulum for which the
length is uniformly altered in time; i. e.,
\begin{equation}
  \label{eq:lut}
  \Le (t ) = \Le_{0} (1 + \varepsilon  t) \, ,
\end{equation}
where $\varepsilon$ is a small parameter with the dimensions of a
reciprocal time.  It will be convenient to let $\tau = \varepsilon t$ and
define a dimensionless time-dependent frequency
\begin{equation}
  \label{eq:Ome}
  \omega (\tau) = \frac{1}{\varepsilon} \sqrt{\frac{g}{\Le(\tau)}} \, .
\end{equation}
Equation (\ref{eq:eqmotnl}) can be thus recast as
\begin{equation}
  \label{eq:linllut}
  \ddot{\vartheta} + 
  2 \frac{\dot{\Le}}{\Le} \, \dot{\vartheta} +
  \omega^{2} (\tau) \, \vartheta = 0 \, ,
\end{equation}
where the dot represents differentiation with respect to $\tau$.  We
limit our analysis to a lengthening pendulum because if the amplitude
of the initial displacement is small, then the resulting displacement
will stay small. On the other hand, if the pendulum is shortening,
then even if the initial displacement is small, the subsequent
displacement will grow in time, violating the linearization
hypothesis.

Using basic properties of Bessel functions~\cite{McLachlan:1955ys},  
the following
\begin{equation}
  \label{eq:sollutin}
  \vartheta (\tau)  =  \frac{1}{\sqrt{1 + \tau}} 
  [ A \;  J_{1} (2 \omega_{0} \sqrt{1 + \tau} ) + 
  B \;  Y_{1} (2 \omega_{0} \sqrt{1 + \tau} ) ]  \, ,
\end{equation}
is a solution to (2.5), with $\omega_{0} = \omega(0)$ and $J_{n} (x)$
and $Y_{n} (x)$ the Bessel functions of $n$th order and first and
second kind, respectively.  The constants $A$ and $B$ must be
determined by the initial conditions, which we take, without loss of
generality, as $ \vartheta (0) = \vartheta_{0}$ and $\dot{\vartheta}
(0) = 0$.  The final result reads as
\begin{eqnarray}
  \label{eq:sollut}
  \vartheta (\tau) & = & 
  \frac{\pi \vartheta_{0} \omega_{0}}{\sqrt{1 +\tau}} \, 
  [  J_{2} (2 \omega_{0}) \, Y_{1} (2 \omega_{0} \sqrt{1 + \tau} )
  \nonumber \\
  &  - &
  Y_{2} (2 \omega_{0}) \, J_{1} (2 \omega_{0} \sqrt{1 + \tau} ) ]  \, .
\end{eqnarray}
As a side comment, we remark that, in spite of its
usual designation, neither Rayleigh nor Lorentz actually examined the
behavior of this pendulum; this was first accomplished much later by
Krutkov and Fock~\cite{Krutkow:1923dq}, who obtained (\ref{eq:sollut})
and also derived Eq.~(\ref{eq:adinv1}) directly therefrom. Indeed,
this solution allows one to investigate in great detail the periods of
this system~\cite{Wickramasinghe:2005yq}.

Since $\varepsilon$ appears in the denominator in the
definition~(\ref{eq:Ome}), $\omega_{0}$ is actually very large (for
example, if $\ell_{0} = 1$~m and $\varepsilon = 0.01$~s$^{-1}$, then
$\omega_{0} \simeq 313$). This suggests to consider the limit
\begin{equation}
  \label{eq:lelut}
  2 \omega_{0} \sqrt{1 + \tau} \gg 1 \, ,
\end{equation}
and then take the leading term in the asymptotic expansion of the
Bessel functions~\cite{McLachlan:1955ys}
\begin{eqnarray}
  \label{eq:asymBess}
  J_{n} (x) & \sim & \sqrt{\frac{2}{\pi x}} 
  \cos \left  ( x - \frac{\pi}{4} - \frac{n \pi}{2} \right ) \, ,
  \nonumber \\
  & & \\
  Y_{n} (x) & \sim & \sqrt{\frac{2}{\pi x}} 
  \sin \left  ( x - \frac{\pi}{4} - \frac{n \pi}{2} \right ) \, .
  \nonumber
\end{eqnarray}
Consequently, we get
\begin{equation}
  \label{eq:sollutaprox}
  \vartheta (\tau) \simeq 
  \frac{\vartheta_{0}}{(1 +\tau)^{3/4}} \, 
  \cos  [ 2 \omega_{0}  ( \sqrt{1 + \tau}  - 1) ] \, .
\end{equation}

By making use of the approximation  (\ref{eq:sollutaprox}), one can
check that the maximum angular amplitude $\vartheta_{\mathrm{max}}$
scales as
\begin{equation}
  \label{eq:thetamax}
  \vartheta_{\mathrm{max}} (\tau) = \frac{\vartheta_{0}}{(1
    +\tau)^{3/4}} = \vartheta_{0} 
  \left [ \frac{\Le_{0}}{\Le (\tau)} \right ]^{3/4} \, ,
\end{equation}
which shows that it is a decreasing function of time.
Moreover, $  \lim_{\tau \rightarrow \infty} \vartheta_{\mathrm{max}} (\tau) = 0$.

\begin{figure}
  \centerline{\includegraphics[width=0.95\columnwidth]{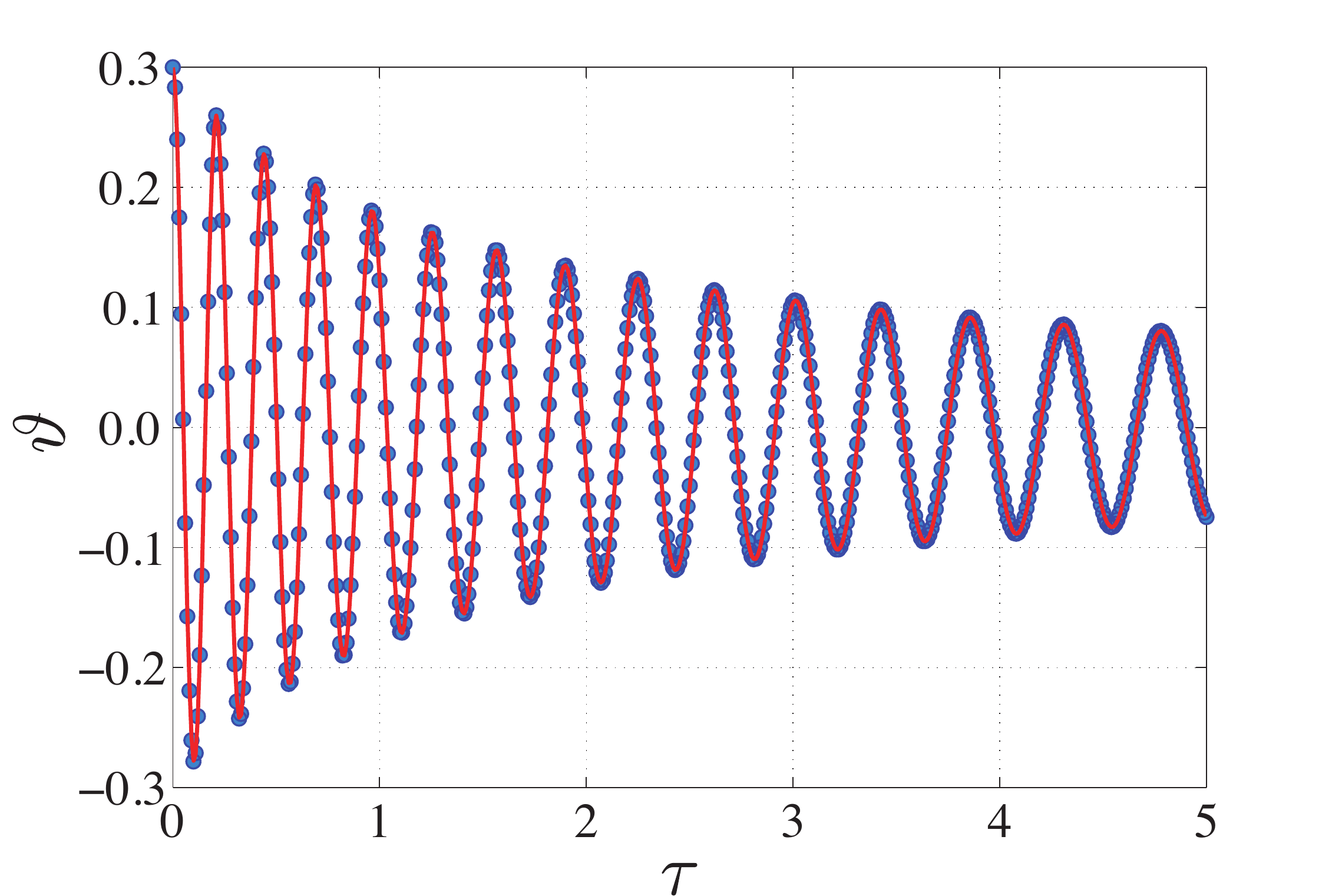}}
  \caption{Exact solution (red continuous line) and asymptotic
    approximation (blue points) for the uniformly varying pendulum
    with $\varepsilon= 0.1$ s$^{-1}$, $\Le_{0} = 1$ m and
    $\vartheta_{0}  = 0.3$ rad.}
\end{figure}

In Fig.~1 both the exact solution (\ref{eq:sollut}) and its asymptotic
approximation (\ref{eq:sollutaprox}) are plotted. The error associated
with (\ref{eq:sollutaprox}) is very small; obviously, this error
is smaller when $\tau$ is larger, since larger $\tau$'s improve the
approximation of (\ref{eq:lelut}). This can be formally expressed as
\begin{equation}
  \label{eq:limformult}
  \lim_{\tau \rightarrow \infty} 2 \omega_{0} 
  \sqrt{1 + \tau} = \infty  \, .
\end{equation}
This limit guarantees that, for a given value of $\omega_{0}$, the
asymptotic expansion is always a good approximation for any $\tau$.
From this perspective we can assert that the validity of the
asymptotic expansion depends on the values of $\vartheta_{0}$ (and
$\varepsilon$) but it is independent of time.

\subsection{Adiabatic invariance}

As pointed out in the Introduction, the concept of adiabatic change is
associated with a variation that occurs infinitely slowly. The
observer who is controlling the changes does not know the internal
state of the system.  In practical terms, this means that the change
is adiabatic when the variation is carried out continuously and so
slowly that the change $\delta \Le$ of the length is very small
compared to the length $\Le$ of the pendulum~\cite{Andrade:1962yq};
i.e.
\begin{equation}
  \label{eq:condad1}
  \frac{\delta \Le}{\Le} \ll 1 \, .
\end{equation}
By considering that the temporal interval in which the variation
$\delta \Le$ is produced coincides with the local period $T$ for the length
$\Le$, we have
\begin{equation}
  \label{eq:nous}
  \delta \Le = T \frac{d\Le}{dt} \, ,
\end{equation}
and recalling that $T = 2 \pi \sqrt{\Le /g}$, we can rewrite
(\ref{eq:condad1}) for the example at hand as
\begin{equation}
  \label{eq:condad2}
  \frac{\delta \Le}{\Le} = \frac{\varepsilon T_{0}}{\sqrt{1 + \tau}} = 
  \frac{2 \pi}{\omega_{0} \sqrt{1 + \tau}} \gg 1   \, ,
\end{equation}
with $T_{0}$ being the period of the pendulum at $\tau = 0$.  This
requirement is independent of $\vartheta_{0}$ and, thus, independent
of the amplitude of the oscillations, which is intuitively expected.

Equation~(\ref{eq:condad2}) clearly suggests that if the change in
length is initially adiabatic, it will remain forever. Moreover, the
adiabatic character of the system will improve as time goes on.

Interestingly enough, condition~(\ref{eq:condad2}) is formally
equivalent to (\ref{eq:lelut}) ensuring the validity of the asymptotic
approximation; they will always be satisfied whenever
\begin{equation}
  \label{eq:condad3}
  \omega_{0} \gg \omega_{\mathrm{lim}} = 2\pi  \, ,
\end{equation}
which, from the definition of $\varepsilon$, can be equivalently
recast as
\begin{equation}
  \label{eq:condad4}
  \varepsilon  \ll \varepsilon_{\mathrm{lim}} = 
 \frac{1}{2 \pi} \sqrt{\frac{g}{\Le_{0}}}  \, .
\end{equation}
Equations (\ref{eq:condad3}) or (\ref{eq:condad4}) (which do not
depend on time) provide a sensible criterion for the adiabatic change
in a uniformly varying pendulum. The lesser the initial length, the
more quickly can be lengthening the pendulum under the adiabatic
hypothesis. Alternatively, $\omega_{\mathrm{lim}}$ can be seen as the
minimum value of $\omega_{0}$ for which the asymptotic expansion is
valid independently of $\tau$.

To give a more quantitative argument, we compute the function
\begin{equation}
  \label{eq:Ehnadin}
  I( \tau) = \frac{H(\tau)}{\nu(\tau)} \, ,
\end{equation}
that turns out to be an adiabatic invariant for arbitrary periodic
motions in one degree of freedom~\cite{Ehrenfest:1916cr}. Here
$H(\tau)$ is the Hamiltonian and $\nu(\tau)$ the frequency of the
oscillations.

For a pendulum, the Hamiltonian is
\begin{equation}
  \label{eq:Hampen}
 H = \frac{1}{2} \frac{p^{2}}{\Le^{2}} + \frac{1}{2} g \Le
 \vartheta^{2} \, ,
\end{equation}
where $p =\Le^{2} d\vartheta/dt$ is the generalized momentum conjugate
to $\vartheta$.

\begin{figure}
  \centerline{\includegraphics[width=0.95\columnwidth]{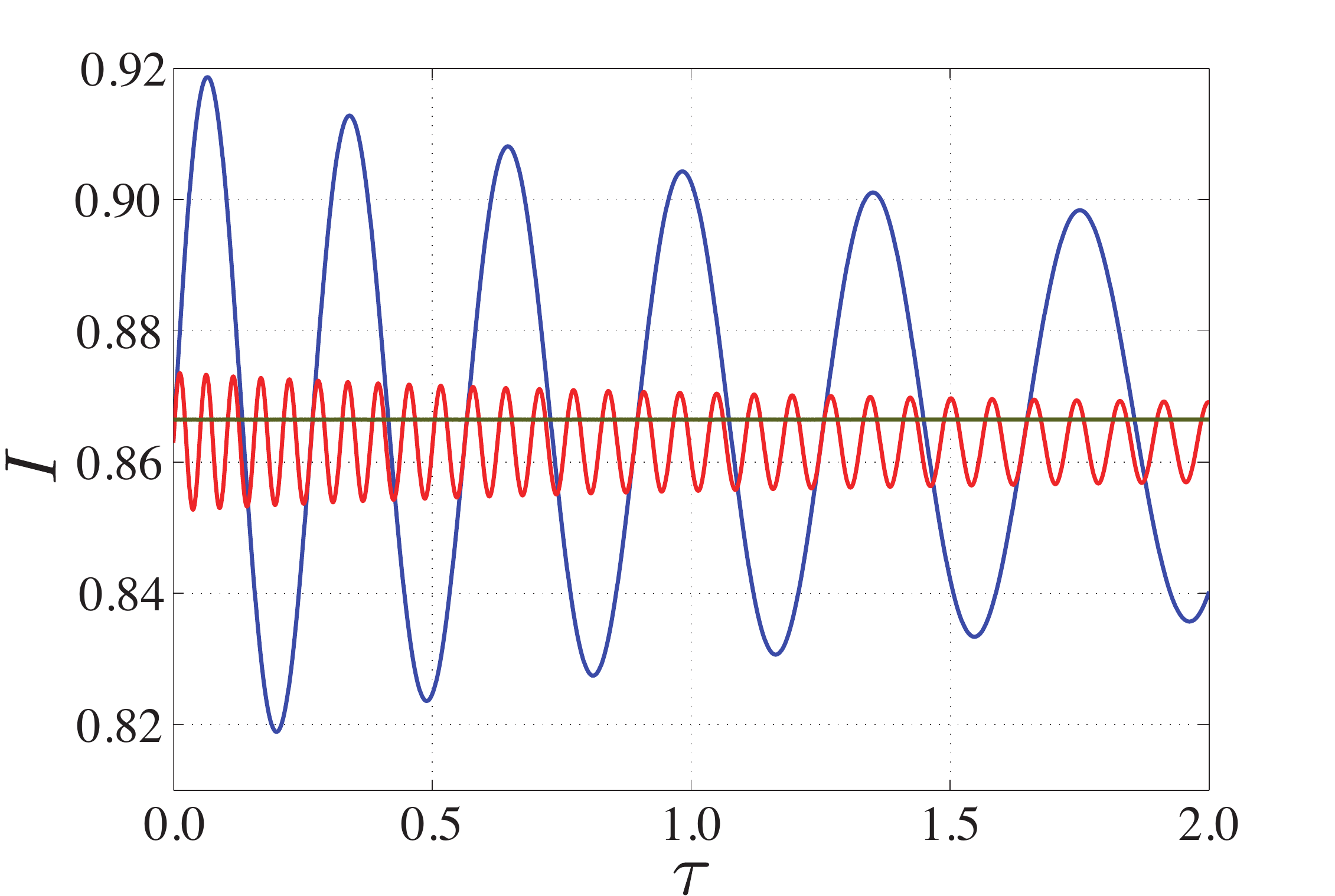}}
  \caption{Plot of $I (\tau)$ as a function of $\tau$ for  
    the uniformly varying pendulum with $\Le_{0} = 1$ m and
    $\vartheta_{0}= 0.3$ rad. The curves corresponds to $\omega_{0} =
    2, 10$ and 1000 (which are associated with the values
    $\varepsilon= 0.2491, 0.0498$ and 0.00005 s$^{-1}$), following the
    decreasing amplitudes.}
\end{figure}

Taking into account the relations~\cite{McLachlan:1955ys}
\begin{equation}
  \label{eq:a3}
  \frac{d}{dx} [x^{-1} J_{1} (x) ] = x^{-1}  J_{2} (x) , 
  \qquad
  \frac{d}{dx} [x^{-1} Y_{1} (x) ] = x^{-1}  Y_{2} (x) \, , 
\end{equation}
and the solution (\ref{eq:sollut}), the angular velocity of the
pendulum can be expressed as
\begin{equation}
  \label{eq:dthetault}
  \dot{\vartheta}  =  \frac{\pi \vartheta_{0} \omega_{0} \varepsilon}{1 +
    \tau}  \, 
  \mathcal{H}_{22} (2 \omega_{0} \sqrt{1 + \tau} ) \, ,
\end{equation}
which immediately leads to
\begin{equation}
  \label{eq:H1lut}
  H ( \tau ) = H_{0} \pi^{2} \omega_{0}^{2}  [  
  \mathcal{H}_{22}^{2} (2  \omega_{0} \sqrt{1 + \tau} )
  + \mathcal{H}_{21}^{2} (2 \omega_{0} \sqrt{1 + \tau} ) ]  \, .
\end{equation}
Here, for notational simplicity, we have introduced the functions
\begin{eqnarray}
  \label{eq:H1}
  \mathcal{H}_{21} ( x ) & = & J_{2} (2 \omega_{0}) \; Y_{1} (x) -
  Y_{2}(2 \omega_{0}) \; J_{1}(x) \, , \nonumber \\ 
  & &\\
  \mathcal{H}_{22} (x) & = & Y_{2} (2 \omega_{0}) \;  J_{2} (x) -
  J_{2} (2 \omega_{0}) \;  Y_{2}(x) \, , \nonumber 
\end{eqnarray}
and $ H_{0} = \vartheta_{0} \omega_{0}^{2} \varepsilon^{2} \Le_{0}^{2}
/2$ is the total energy at $\tau = 0$. Finally, since $\nu (\tau) =
\nu_{0}/\sqrt{1 + \tau}$, with $v_{0} = \omega_{0} \varepsilon/2 \pi$,
we get
\begin{eqnarray}
  \label{eq:I1lut}
  I (\tau) & = & \frac{H_{0 }}{\nu_{0}}  \sqrt{1+ \tau} \;
  \pi^{2} \omega_{0}^{2} \nonumber \\
  &  \times & [  \mathcal{H}_{22}^{2} (2  \omega_{0} \sqrt{1 + \tau} )
  + \mathcal{H}_{21}^{2} (2 \omega_{0} \sqrt{1 + \tau} ) ]  \, .
\end{eqnarray}
Because $I (\tau)$ is a time-dependent function, it will not be, in
general, an adiabatic invariant. In other words, for arbitrary values
of the parameters, the solution (\ref{eq:sollut}) will not be
associated with adiabatic changes in the length of the pendulum.  The
function $I(\tau)$ is shown in Fig.~2 for different values of
$\omega_{0}$. One immediately concludes that the larger $\omega_{0}$,
the lesser time-dependent $I (\tau)$ becomes, which is in full
agreement with~(\ref{eq:condad3}).

To complete the analysis, we proceed to calculate $I (\tau)$ with the
asymptotic approximations for the Bessel functions as in
Eq.~(\ref{eq:sollutaprox}). After a direct manipulation, we end up
with
\begin{equation}
  \label{eq:Haplut}
  H (\tau ) \simeq \frac{H_{0}}{\sqrt{1 + \tau}}
\end{equation}
so that
\begin{equation}
  \label{eq:H0aplut}
  I (\tau ) \simeq \frac{H_{0}}{\nu_{0}} = \pi \Le_{0}^{3/2}
  \vartheta_{0}^{2} g^{1/2} \, .
\end{equation}
This is an important result: for large enough values of $\tau$, the
quantity $I(\tau)$ becomes an adiabatic invariant. This validates the
previously suggested conclusion: the condition establishing the validity
of the asymptotic expansion of the Bessel functions is conceptually
equivalent to the condition of adiabatic invariance.

\section{The exponentially varying pendulum}

\subsection{Basic equations of motion}

We turn now our attention to the instance where the length of the pendulum
is altered non uniformly.  More concretely, we take
\begin{equation}
  \label{eq:nlut}
  \Le (t ) = \Le_{0}  \, e^{\varepsilon t}  \, .
\end{equation}
Equation~(\ref{eq:eqmotnl}), when small oscillations are considered,
reduces in this case to
\begin{equation}
  \label{eq:linlnlut}
  \ddot{\vartheta} + 
  2 \dot{\vartheta} +
  \omega_{0}^{2} e^{- \tau} \vartheta = 0 \, .
\end{equation}
The change $\vartheta = \theta e^{- \tau}$ gives
\begin{equation}
  \ddot{\theta} + 
  (\omega_{0}^{2} e^{- \tau} - 1) \theta = 0 \, ,
\end{equation}
which has again an exact solution in terms of Bessel
functions~\cite{Kamke:1974mz}. The result, employing the
original variables, reads as
\begin{equation}
  \label{eq:solnlutpre}
  \vartheta (\tau)  =    e^{- \tau} \, 
  [  A \;  J_{2} (2 \omega_{0} e^{- \tau/2} ) +  B  \; Y_{2} (2 \omega_{0} e^{- \tau /2} ) ] \, .
\end{equation}
To fix the constants $A$ and $B$ we take the same initial conditions
as before, namely $ \vartheta (0) = \vartheta_{0}$ and
$\dot{\vartheta} (0) = 0$.  Applying the relations
\begin{equation}
  \label{eq:3}
  \frac{d}{dx} [x^{2} J_{2} (x) ] = x^{2}  J_{1} (x) , 
\qquad
  \frac{d}{dx} [x^{2} Y_{2} (x) ] = x^{2}  Y_{1} (x) 
\end{equation}
in conjunction with the change of variable $x= 2 \omega_{0} 
e^{-  \tau/2}$ and the Wronskian
\begin{equation}
  \label{eq:4}
  J_{1} (x) Y_{2} (x) - J_{2} (x) Y_{1} (x) = - \frac{2}{\pi x} \, ,
\end{equation}
one can show that the  final solution  is
\begin{eqnarray}
  \label{eq:solnlut}
  \vartheta (\tau) & = & 
  \pi \vartheta_{0} \omega_{0} e^{- \tau} \, 
  [  Y_{1} (2 \omega_{0}) \;  J_{2} (2 \omega_{0} e^{- \tau/2} )
  \nonumber \\
  & - &
  J_{1} (2 \omega_{0}) \; Y_{2} (2 \omega_{0} e^{- \tau /2} ) ] \, .
\end{eqnarray}

Much in the same way as for  the pendulum with
uniformly varying length, if 
\begin{equation}
  \label{eq:lenlut}
  2 \omega_{0} e^{ - \tau /2} \gg 1 \, ,
\end{equation}
is satisfied, we can replace equation~(\ref{eq:solnlut}) by its
asymptotic approximation, leading to
\begin{equation}
  \label{eq:solnlutapp}
  \vartheta (\tau) \simeq
  \vartheta_{0}  e^{- 3 \tau/4 } \, 
  \cos [   2 \omega_{0} ( 1 - e^{- \tau/2} )  ] \, .
\end{equation}
Within this approximation the maximum angular amplitude
$\vartheta_{\mathrm{max}}$ scales exactly as in
Eq.~(\ref{eq:thetamax}), and also $ \lim_{\tau \rightarrow \infty}
\vartheta_{\mathrm{max}} (\tau) = 0$. However, as expected, the
angular amplitude $\vartheta_{\mathrm{max}}$ falls off more quickly in
this case. This behavior can be clearly appreciated in Fig.~3, where
both the exact solution (\ref{eq:solnlut}) and its asymptotic
approximation (\ref{eq:solnlutapp}) are plotted. The agreement is
again remarkable.

\begin{figure}[t]
  \centerline{\includegraphics[width=0.95\columnwidth]{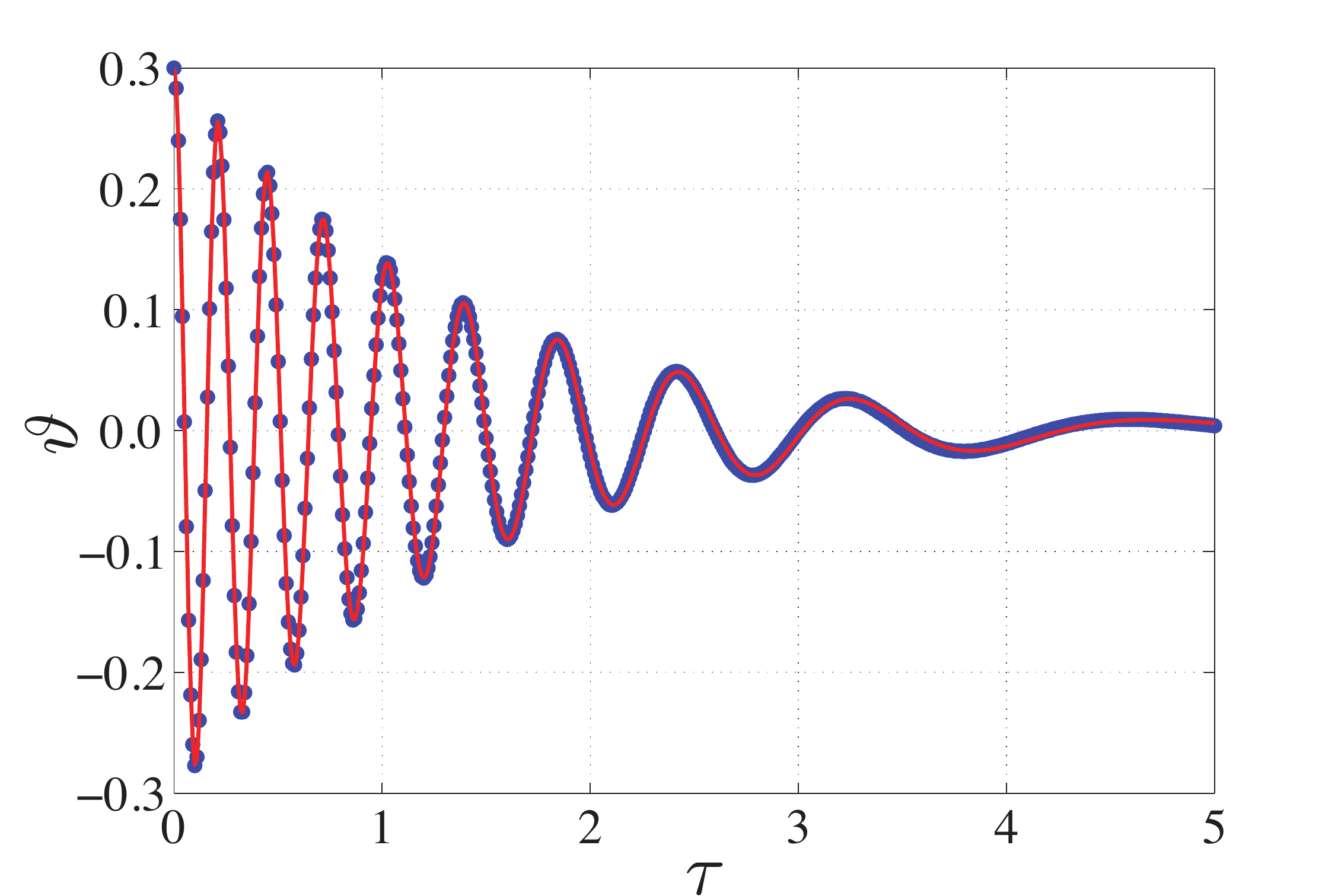}}
  \caption{Plot of the exact solution (in red) and the asymptotic
    approximation (blue points) for the exponentially varying pendulum
   with the same parameters as in Fig.~1.}
\end{figure}

In contradistinction with the situation described by
Eq.~(\ref{eq:limformult}), the exponentially varying pendulum leads to
the limit relation
\begin{equation}
  \label{eq:limformunlt}
  \lim_{\tau \rightarrow \infty} 2 \omega_{0} e^{- \tau/2} = 0 \, .
\end{equation}
This points out the more important conceptual difference between these
two cases: for the uniformly varying pendulum the validity of the
asymptotic expansion of the Bessel functions only depends on the
values of $\varepsilon$ and $\vartheta_{0}$, but it is independent on
the time. For the exponentially varying pendulum the validity of that
approximation is time dependent and for large values of the time this
approximation breaks down.

At first glance, this important difference between the two penduli can
seems only a formal one. However, as we shall see in the following
it has strong implications for the adiabatic invariance.

\subsection{Adiabatic invariance}

We next analyze the adiabatic change for this
example. Equation~(\ref{eq:nous}) applied to (\ref{eq:nlut}) gives as
a requirement for adiabatic invariance
\begin{equation}
  \label{eq:condd21}
  \frac{\delta \Le}{\Le} =  \varepsilon T_{0}  e^{\tau /2} = 
\frac{2 \pi}{\omega_{0}} e^{\tau /2} \ll  1  \, ,
\end{equation}
which  again is  formally equivalent to the condition
(\ref{eq:lenlut}) for the validity of the asymptotic approximation of
the Bessel functions.

This indicates that even if the change of length is initially
adiabatic, it will remain so only for a finite interval of time.  To put
in  another way, (\ref{eq:condd21})  holds true whenever
\begin{equation}
  \label{eq:condd2b}
  \omega_{0} \gg  e^{\tau/2} \omega_{\mathrm{lim}} = 2\pi  e^{\tau / 2} \, ,
\end{equation}
or, recalling the definition of $\omega (\tau)$, 
\begin{equation}
  \label{eq:condad3b}
  \tau \ll \tau_{\mathrm{lim}} = 2
\ln \left ( \frac{1}{\varepsilon T_{0}} \right ) \, .
\end{equation}
Accordingly, there does not exist a minimum fixed value of
$\omega_{0}$ such that adiabaticity holds true forever. In our opinion
this is the more important lesson from this paper: the adiabatic
condition does not need to hold, in general, for all times.

The formal equivalence discussed so far provides a mathematical
interpretation fo (\ref{eq:condad3b}): the change will be adiabatic in
those conditions in which the asymptotic expansion of Bessel functions
is justified.

Note that the change of the length in this example is infinitely
continuously differentiable, but the adiabaticity holds true only in a
time interval fixed by the values of $\Le_{0}$ and $\varepsilon$. This
by no means contradict Arnold's adiabaticity requirement mentioned in
the Introduction (that is, that the change of the length must be twice
continuously differentiable), since Arnold explicitly considers a
finite time interval in the definition of the adiabatic
invariants~\cite{Arnold:1978yq}.

\begin{figure}[t]
  \centerline{\includegraphics[width=0.95\columnwidth]{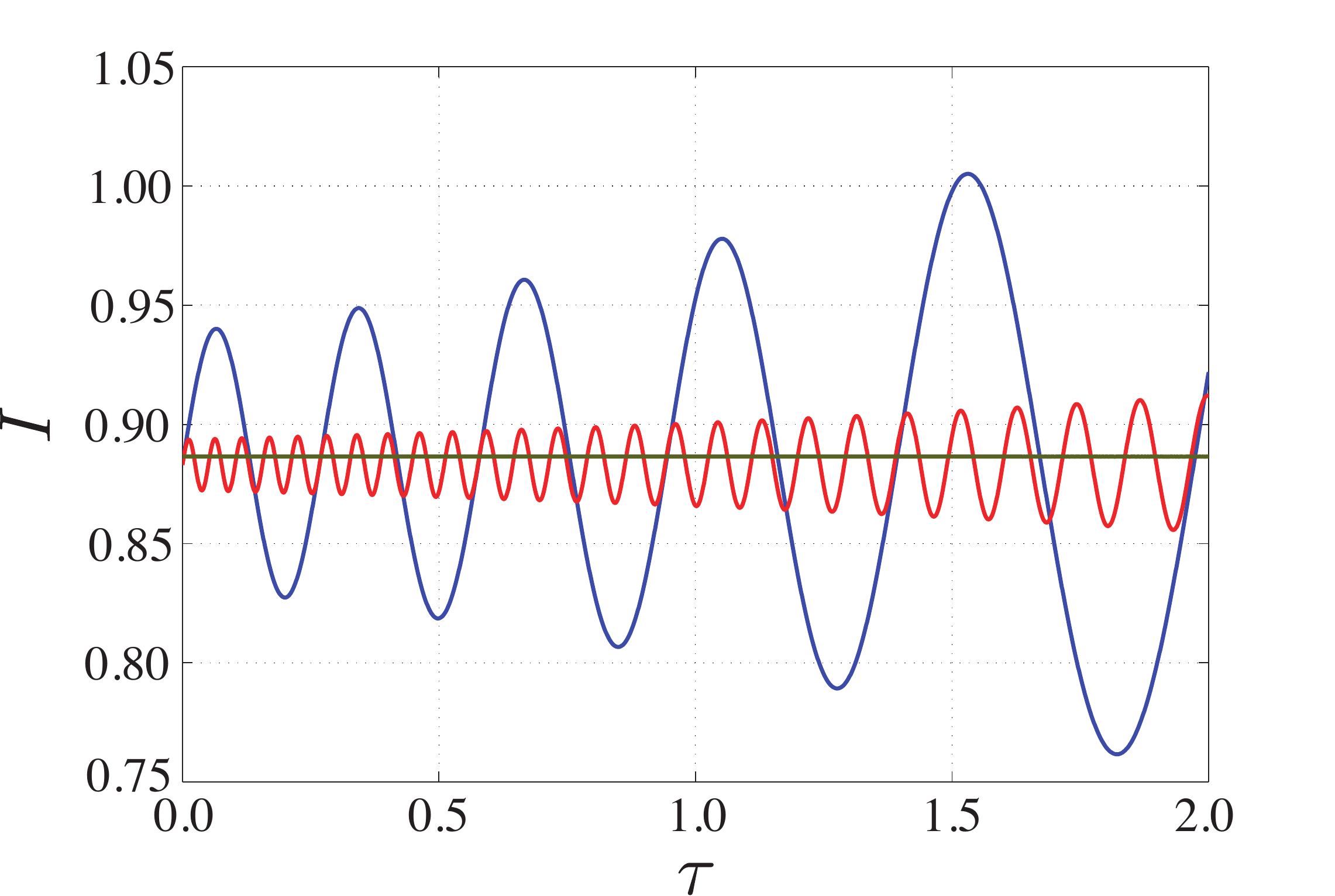}}
  \caption{ Plot of $I (\tau)$ as a function of $\tau$ for  
    the exponentially  varying pendulum with $\Le_{0} = 1$ m and
    $\vartheta_{0}= 0.3$ rad. The curves corresponds to $\omega_{0} =
    2, 10$ and 1000 (which are associated with the values
    $\varepsilon= 0.2491, 0.0498$ and 0.00005 s$^{-1}$), following the
    decreasing amplitudes.} 
\end{figure}

Finally, we calculate explicitly the total energy for this
case. Using again the Hamiltonian (\ref{eq:Hampen}) and the solution
(\ref{eq:solnlut}) and its time derivative, we get
\begin{eqnarray}
  \label{eq:H1nlut}
  H ( \tau ) & = & H_{0 } \;   \pi^{2} \omega_{0}^{2}  e^{- \tau} \nonumber \\
  & \times & [  
  \mathcal{H}_{11}^{2} (2  \omega_{0} e^{ - \tau/2}) 
  + \mathcal{H}_{12}^{2} (2 \omega_{0} e^{- \tau /2})  ]  \, ,
\end{eqnarray}
with
\begin{eqnarray}
  \label{eq:H2}
  \mathcal{H}_{12} ( x ) & = & Y_{1} (2 \omega_{0}) \;  J_ {2} ( x )  -  
 J_{1} (2 \omega_{0}) \; Y_{2} ( x ) \, , \nonumber \\ 
  & &\\
  \mathcal{H}_{11} ( x ) & = & J_{1} (2 \omega_{0})  \; Y_{1} ( x ) -
  Y_{1} (2 \omega_{0}) \; J_{1} ( x ) \, . \nonumber 
\end{eqnarray}
In consequence, $I (\tau)$ becomes
\begin{equation}
  \label{eq:I1nlut}
  I (\tau)  =  \frac{H_{0 }}{\nu_{0}}  \, 
  \pi^{2} \omega_{0}^{2} e^{-  \tau /2}  
  [  \mathcal{H}_{11}^{2} (2  \omega_{0} e^{- \tau/2}) 
  + \mathcal{H}_{12}^{2} (2 \omega_{0} e^{- \tau/2} ) ]  \, .
\end{equation}

The function $I(\tau)$ is represented in Fig.~4. As we can see, the
fluctuations of $I(\tau)$ increase with time. Thus, for large enough
time, $I (\tau)$ will never be an invariant quantity, irrespective of
the value of $\omega_{0}$. However, for the time window chosen in the
figure, we see that for $\omega_{0} = 1000$, $I (\tau)$ looks
invariant over the entire interval.

Our last step is to calculate $I (\tau)$ using the asymptotic
approximations for the Bessel functions. Now, we have
\begin{equation}
  H (\tau ) \simeq  H_{0} e^{- \tau /2} \, 
\end{equation}
so that
\begin{equation}
  \label{eq:H0aplut2}
  I (\tau ) \simeq \frac{H_{0}}{\nu_{0}} = \pi \Le_{0}^{3/2}
  \vartheta_{0}^{2} g^{1/2} \, ,
\end{equation}
which is identical to what we have obtained for the uniformly varying
pendulum.

\section{Concluding remarks}

We have explored in detail two nontrivial yet solvable examples of
penduli of varying length with the purpose of a better understanding
of the concept of adiabatic invariance.

The ambiguous criteria of ``infinitely slow variation'', or the
assumption of ``ignorance of the internal state of the system on the
part of the person controlling the variable parameter'', usually
employed to establish the condition of adiabatic change, are replaced
here by more quantitative criteria.

For the two penduli considered in this paper, we have shown that the
physical meaning of adiabatic change is formally contained in the
mathematical condition of validity for the asymptotic expansion of the
Bessel functions: the validity of the asymptotic approximation implies
adiabatic change and vice versa.

The analysis carried out in this work invites to a more general
reflection: it is important to pay special attention to the meaning of
mathematical approximations. Actually, the radius of convergence of
some systematic approximation to an exact solution has always a
physical origin.

\acknowledgments

The original ideas in this paper originated from a long cooperation
with the late Richard Barakat. Over the years, they have been further
developed and completed with questions, suggestions, criticism, and
advice from many students and colleagues. Particular thanks for help
in various ways go to E. Bernab\'eu, J. F. Cari\~{n}ena, A. Galindo,
H. de Guise, H. Kastrup,  A. B. Klimov, G. Leuchs and J. J. Monz\'on.

We are indebted to two anonymous referees for valuable comments. 

This paper is dedicated to the memory of coauthor J. Zoido, who
unexpectedly passed away during the preparation of the final version.

This work is partially supported by the Spanish DGI (Grants FIS2008-04356 and
FIS2011-26786) and the UCM-BSCH program (Grant GR-920992).


\end{document}